\documentclass[fleqn,twoside]{article}%
\usepackage{amsmath}
\usepackage{amsfonts}
\usepackage{amssymb}
\usepackage{graphicx}
\usepackage{cite}
\def\be{\begin{equation}}
\def\ee{\end{equation}}
\def\bi{\bibitem}
\begin{document}
\title{Hydrodynamic self-similar cosmological models.}
\author{Abhik Kumar Sanyal$^1$, A. Banerjee$^2$ and D. Ray$^3$}
\maketitle
\noindent
\begin{center}
\noindent
$^1, ^2$ Dept of Physics, Jadavpur University,\\
Calcutta-700032, India.\\
$^3$ Dept of Mathematics, Jadavpur University,\\
Calcutta-700032, India.\\
\end{center}
\footnotetext{\noindent
Electronic address:\\
\noindent
$^1$ sanyal\_ ak@yahoo.com \\
Present address: Dept. of Physics, Jangipur College, India - 742213.}
\noindent
\abstract{Hydrodynamic self-similar solutions, as obtained by Chi [J. Math. Phys. 24, 2532 (1983)] have been generalized by introducing new variables in place of the old space and time variables. A systematic procedure of obtaining a complete set of solutions has been suggested. The Newtonian analogs of all homogeneous isotropic Friedmann dust universes with spatial curvature $k = 0, \pm 1$ have been given.}\\
\maketitle
\flushbottom

\section{Introduction:}

In hydrodynamics the idea of self-similarity has been exploited with considerable success to simplify time-dependent problems. Self-similar solutions are often the leading terms in an asymptotic expansion of a non-self-similar evolution, in a regime far from the initial conditions and far from the influence of boundary conditions. The behaviors of the self-similar solutions are likely to be encountered in nature and as such they have greater physical interest than merely being a special class of mathematical solutions. This concerns the concept of intermediate asymptotic as reviewed by Barenblatt and Zeldovich \cite{1}.\\

In a pioneering work, Henriksen and Wesson \cite{2} have discussed some Newtonian and relativistic self-similar cosmological models. They have presented some solutions by considering two constants of nature, viz., the Newtonian gravitational constant `$G$' and a characteristic velocity. Chi \cite{3} thereafter extended their work giving a set of new solutions considering `$G$' as the only constant in nature. One of the interesting solutions obtained by Chi \cite{3} is the Newtonian analog of the Einstein-de Sitter cosmological model with vanishing spatial curvature.\\

In the present paper, while exploring the complete set of hydrodynamic self-similar cosmological model solutions, we find the Newtonian analogs of all homogeneous isotropic Friedmann \cite{4} dust universes with spatial curvature $k = 0~\mathrm{and}~\pm 1$. The complete set of hydrodynamic equations that governs spherically symmetric, isentropic fluid flow in an internal gravitational field includes the continuity equation, the momentum equation, the energy equation, and the gravitational equation. These equations are:

\be \label{1} {\partial \rho\over \partial t} + v {\partial \rho\over \partial r} = -{\rho\over r^2} {\partial\over \partial r}(r^2v),\ee
\be \label{2} {\partial v\over \partial t} + v {\partial v\over \partial r} = g - {1\over \rho}\left({\partial \rho\over \partial r}\right),\ee
\be \label{3} {\partial\over \partial t}(p\rho^{-\gamma}) + v {\partial\over \partial r}(p\rho^{-\gamma}) = 0,\ee
\be \label{4} {\partial g\over \partial t} + v {\partial g\over \partial r} = - {2vg\over r},\ee
where, $\rho$, $p$ and $v$ are, respectively, the local density, pressure, and velocity; $g$ is the localized gravitational acceleration; and $\gamma$ is the polytropic index.\\

Now, a one-parameter self-similar solution (which represents the intrinsic evolution behavior of a system not dependent on the incidental details of particular initial or boundary conditions) is sought for the density, velocity, pressure, and gravitational acceleration. If the Newtonian gravitational constant `$G$' is assumed to be the only constant in nature and further if there is no characteristic velocity, one can then introduce the dimensionless quantities $\eta$, $V$, $\phi$, and $Q$ to obtain the general forms of the physical quantities as follows:

\be\label{5}\begin{split}& \rho = \left({\lambda\over r^{\theta}}\right)\eta(\xi),\hspace{1.2 in} v = \left({r\over t}\right)V(\xi),\\& g = \left({r\over t^2}\right)\phi(\xi),\hspace{1.24 in} p = \left({\lambda\over r^{\theta-2} t^2}\right)Q(\xi),\end{split}\ee
with
\be \label{6}\xi = {G\lambda t^2 \over r^\theta}.\ee
In the above $\theta$ is a constant of the order unity. The parameter $\lambda$ is introduced to make $\xi$ dimensionless.\\

Now Eqs. \eqref{1}-\eqref{4} can be simplified in the following manner without assuming self-similarity. Since $r$ and $t$ are independent variables, Eq. \eqref{1} can be rewritten as:
\be {\partial \over \partial t}(r^2\rho) + {\partial \over \partial r}(r^2\rho v) = 0,\ee
and so, there must exist a function $\psi$, such that,

\be\label{7} r^2\rho = {\partial \psi \over \partial r}, \hspace{1.4 in} r^2\rho v = - {\partial \psi\over \partial t}.\ee
The relations \eqref{7} lead us to write the equation,

\be\label{8} {\partial \psi\over \partial t} + v {\partial \psi\over \partial r} = 0.\ee
Comparing Eq. \eqref{3} with Eq. \eqref{8} it is possible to obtain the relation:

\be \label{9} {{\partial \psi/\partial t}\over {\partial \psi/ \partial r}} = -v = {{\partial (p\rho^{-\gamma})/ \partial t}\over {\partial (p\rho^{-\gamma})/ \partial r}}, \ee
which implies that the quantity $(p\rho^{-\gamma})$ must be a function of $\psi$. We therefore write,

\be\label{10} p\rho^{-\gamma} = f(\psi).\ee
Further in view of \eqref{9} one can express the velocity $v$ as,

\be \label{11} v = {\partial r\over \partial t}\Big |_{\psi = \mathrm{constant}}.\ee
Equation \eqref{11} implies that the new variable $\psi$ characterizes the individual fluid particles.\\

In the following sections we have given a number of solutions of the set of hydrodynamic equations in the presence of gravity. These equations are written in terms of new
variables $\psi$ and $t$ in place of $r$ and $t$, so that on introducing the self-similar variables it is possible to get more general solutions in different cases. The technique followed in this paper leads us to a systematic derivation of different solutions, instead of obtaining only a few as special cases in an ad hoc manner from the equations given by Chi \cite{3}. It is to be noted that the solutions are finally expressed in terms of the old variables $r$ and $t$.\\

In Section 2 we introduce the new variables in place of the old space and time variables $r$ and $t$ in the hydrodynamic equations, without introducing the assumption of self-similarity at this stage. Finally, in Section 3 the self-similar variables are introduced in the equations. Explicit solutions in different cases are given in Sections 4 and 5. Some of these are identical to those given by Chi \cite{3}. In Section 5, we discuss the isotropic homogeneous dust universes with spatial curvature $k = 0, \pm 1$.

\section{Change of variables:}

In this section we shall change the set of independent variables from $(r, t)$ to $(\psi, t)$ in the following manner:

\be \label{21}\begin{split}& {\partial \over \partial t}\Big|_{\mathrm{old}} = {\partial \over \partial t}\Big|_{\mathrm{new}} + {\partial\psi \over \partial t}{\partial \over \partial \psi},\\&
{\partial \over \partial r}\Big|_{\mathrm{old}} = {1\over r_{\psi}}{\partial \over \partial \psi},\end{split}\ee
where, ${\partial \over \partial t}\Big|_{\mathrm{old}}$ implies a time derivative keeping $r$ constant, ${\partial \over \partial t}\Big|_{\mathrm{new}}$ implies a time derivative keeping $\psi$ constant, and ${\partial \over \partial r}\Big|_{\mathrm{old}}$ implies a derivative with respect to the radial vector $r$ keeping $t$ constant. So,

\be\label{20} {\partial v\over \partial t}\Big|_{\mathrm{old}} = v_{t_{\mathrm{new}}} + v_{\psi}{\partial\psi \over \partial t};\hspace{1.2 in} {\partial v\over \partial r}\Big|_{\mathrm{old}} = v_{\psi} {\partial \psi \over \partial r},\ee
where, $v_{t_{\mathrm{new}}} = {\partial v\over \partial t}\Big|_{\mathrm{new}}$, i.e. at constant $\psi$, also $v_{\psi}  = {\partial v\over \partial t}$. Combining these two equations \eqref{20} with Eq. \eqref{8} we obtain quite naturally (from now on we shall refer to $v_{t_{\mathrm{new}}}$ at constant $\psi$ as $v_t$) the relation,

\be \label{22} {\partial v\over \partial t}\Big|_{\mathrm{old}} + v {\partial v\over \partial r}\Big|_{\mathrm{old}} = v_t.\ee
Again,

\be \label{23} {\partial \rho\over \partial r}\Big|_{\mathrm{old}} = {p_\psi\over r_\psi}.\ee
So combining \eqref{2}, \eqref{22} and \eqref{23}, we obtain,

\be \label{24} r_{tt} = {1\over \rho}\left({p_\psi\over r_\psi}\right) - g = 0,\ee
where, we have written $r_t$ for ${\partial r\over \partial t}\Big|_\psi$. Again, by the above transformation rule given by equation \eqref{21}, we further obtain,

\be {\partial g\over \partial t}\Big|_{\mathrm{old}} = g_{t_{\mathrm{new}}} + g_{\psi}{\partial \psi\over \partial t},\ee
where, $g_{t_{\mathrm{new}}}$ is ${\partial g\over \partial t}$ at constant $\psi$. Also, ${\partial g\over\partial r}\Big|_{\mathrm{old}} = g_{\psi_{\mathrm{new}}}\times {\partial \psi\over\partial r}$, where $g_{\psi_{\mathrm{new}}} = {\partial g\over \partial \psi}$. Using these relations and \eqref{8}, we get,

\be {\partial g\over\partial t}\Big|_{\mathrm{old}} + v{\partial g\over\partial r}\Big|_{\mathrm{old}} = g_t.\ee
Here and in what follows we write only $g$, for $({\partial g\over \partial t})\Big|_{\psi}$, and similarly for other variables. The above equation together with Eqs. \eqref{4} and \eqref{11} yields,

\be {g_t\over g} + 2{r_t\over r} = 0,\ee
which on integration, gives us the relation,

\be \label{25} gr^2 = C(\psi).\ee
Equation \eqref{7} further implies that

\be\label{26} \rho = {1\over r^2 r_{\psi}}.\ee
Using the relation \eqref{25}, Eq. \eqref{24} can be written as,

\be\label{27} r_{tt} + {1\over \rho}\left({\partial \rho\over \partial r}\right) - {C\over r^2} =0.\ee
We make it clear at this point that the use of new variables, that is $(\psi, t)$ instead of the old $(r, t)$, will help us to derive many solutions systematically. But sometimes we use the old variables as well, for convenience. To find the pressure, for example, we use Eq. \eqref{27}, where $p$ is written as a function of the old variable $r$. Finally, of course, all the quantities have to be obtained as functions of $r$ and $t$.

\section{Self-similar Solutions:}

In view of equations \eqref{5}, \eqref{6}, \eqref{9}, and \eqref{25} we readily obtain the following relations:

\be \label{31} p\rho^{-\gamma} = f(\psi) = G\lambda^{(2-\gamma)} r^{\{\theta(\gamma-2)+2\}}\eta^{-\gamma}Q\xi^{-1},\ee
and,
\be\label{32} gr^2 = C(\psi) = G\lambda^{(3-\theta)}\phi\xi^{-1}.\ee
One can conclude from \eqref{31} that unless $\theta(\gamma-2)+2 = 0$, the expression for the spatial coordinate $r$ must be in the form of a product of a function of $\psi$ and a function of $\xi$. The same conclusion can be made without any loss of generality from the relation \eqref{32} unless $(3 - \theta) = 0$. These results lead us to consider Eqs. \eqref{31} and \eqref{32} only for two distinct cases, viz., $\theta = 3$ and $\theta \ne 3$. Again for the case $\theta = 3$, it is found from the Eq. \eqref{32} that, if $C(\psi)$ is not a constant, $\xi$ must be a function of $\psi$. This in turn yields, in view of \eqref{31}, the result $\theta(\gamma-2)+2 = 0$ or the $\gamma = {4\over 3}$ polytropic relation.

\section{Solutions for $\theta = 3$:}
\textbf{A. Case 1}. $\theta = 3$, $C(\psi) \ne \mathrm{constant}$, i.e. $\xi = \xi(\psi),~\gamma = {4\over 3}:$\\

In view of Eqs. \eqref{5} and \eqref{11} we find.

\be \label{41} v = r_{t} = \left({r\over t}\right) V.\ee
Finding $r_t$ from Eq. \eqref{6} and substituting its value in the above equation \eqref{41} one finds

\be \label{42} V = {2\over 3}.\ee
Let us consider here a special case for the matter content, that is, for dust with $p = 0$. Using in this case Eq. \eqref{42} in \eqref{41} one can calculate $r_{tt}$ as

\be r_{tt} = -{2\over 9}\left({r\over t^2}\right),\ee
which, when substituting in Eq. \eqref{27}, gives

\be \label{43} C(\psi) = -{2\over 9}\left({G\lambda\over \xi}\right).\ee
The above expression for $C(\psi)$ being substituted in \eqref{32} immediately yields

\be \label{44} \phi = -{2\over 9}.\ee
Again Eqs. \eqref{5} and \eqref{26} are combined to give

\be \rho = {1\over r^2 r_{\psi}} = {\lambda \eta\over r^3}.\ee
Computing $r_\psi$, from Eq. \eqref{6} and using it in the above equation
we finally obtain the relation

\be \label{45} \eta = -\left({3\over \lambda}\right) \xi \psi_{\xi}.\ee
So ultimately with the values of $V$ and $\phi$ given by Eqs. \eqref{42} and \eqref{44}, it is quite easy to find $v$ and $g$ from Eq. \eqref{5}. Also since $\eta$ is given as a function of $\xi$ by Eq. \eqref{45}, it is possible to find $\rho$ from Eq. \eqref{5} as a function of $\xi$, where $\xi$ is given by Eq. \eqref{6}. Thus the solutions are obtained for the dust case $p = 0$. It is interesting to note that for the special choice of $\theta = 3$ and $C(\psi) \ne$constant, the numerical values of the parameters $\gamma$ and $V$ are uniquely determined, viz. $\gamma = {4\over 3}$ and $V = {2\over 3}$.\\

\noindent
\textbf{B. Case 2}. $\theta = 3$, $C(\psi) = \mathrm{constant:}$\\

In this case $g$ is a function of $r$ only, as can be quite easily observed from Eq. \eqref{25}. Also from Eq. \eqref{32} it is found that

\be\label{46} \phi = {C\over G\lambda}\xi.\ee
In view of Eqs. \eqref{5}, \eqref{11}, and \eqref{26} we obtain the following two relations

\be r_{\psi} = {r\over \lambda \eta};~~~~~~~r_t = \left({r\over t}\right)V,\ee
which in turn yield

\be -{\partial \psi\over \partial t}\Big|_{r = \mathrm{constant}} = {{\lambda\eta V\over t}}.\ee
The above relation can be written in a useful form like

\be {\partial \psi\over \partial({G\lambda t^2\over r^2})}\Big|_{r = \mathrm{constant}} = {\partial \psi\over \partial \xi}\Big|_{r = \mathrm{constant}} = -{\lambda\eta V\over 2\xi},\ee
the integration of which can be immediately be performed to yield

\be\label{47} \psi = - \int{\lambda\eta V\over 2\xi}d\xi + b(r) = a(\xi) + b(r),\ee
where $b(r)$ is an arbitrary function of $r$. We should note that $\xi(\psi, t)$ and $r(\psi, t)$ are related to each other in the form given in \eqref{47}. From Eq. \eqref{47} one can obtain

\be {d a(\xi)\over d\xi}(\xi){\partial\xi\over \partial t}\Big|_{\psi = \mathrm{constant}} + {db(r)\over dr}\left({\partial r\over \partial t}\right)\Big|_{\psi = \mathrm{constant}} = 0,\ee
Using the definition of $\xi$ given in \eqref{6} and \eqref{11} in the above equation, we get for $V \ne 0$

\be \xi {da(\xi)\over d\xi}\left({2\over V}-3\right) = -r \left({db(r)\over dr}\right).\ee
Since the left-hand side of the above equation is a function of $\xi$, while the right-hand side is a function of $r$, the above equation holds in view of \eqref{6} only if both the sides are separately equal to a constant, say ($-D$). Then one obtains

\be\label{48} r \left({db(r)\over dr}\right) = D = \xi {da(\xi)\over d\xi}\left(3 -{2\over V}\right).\ee
The relation \eqref{48} on integration yields

\be b(r) = D\ln{r} + \mathrm{constant},\ee
while Eq. \eqref{47} takes the form

\be \label{49} \psi = a(\xi) + D\ln{r} + \mathrm{constant}.\ee
Equation \eqref{49} has been obtained with the only assumption
that $V \ne 0$ . On the contrary, if we consider $V = {2\over 3}$, then from Eq. \eqref{48} we get
$D = 0$ and as such $\psi = \psi(\xi)$, which has already been considered above in case 1. So in the following section, we discuss cases for $V \ne {2\over3}$. Let us also consider dust, in the form for which $p = 0$. Equations \eqref{24} and \eqref{22} together now yield

\be {\partial v \over \partial t}\Big|_{r = \mathrm{constant}} + v{\partial v\over \partial r}\Big|_{r = \mathrm{constant}} = g.\ee
Finding explicit expressions for ${\partial v\over \partial t}\Big|_r$ and $v{\partial v\over \partial r}\Big|_t$ from Eq. \eqref{5} and substituting these expressions together with that of $g$ from Eq. \eqref{25}, in the above equation, we readily obtain the following relation

\be \xi {dV\over d\xi}(2-3V) + V^2 - V = {C\over G\lambda}\xi,\ee
which admits the solution

\be \label{410} V = \sqrt{\xi},~~~~~{C\over G\lambda} = -{1\over 2}.\ee
Using Eqs. \eqref{5}, \eqref{32}, and \eqref{410} we arrive at the following pair of relations

\be \label{411} v = \sqrt{\xi}\left({r\over t}\right),\ee
and
\be \label{412} \phi = -{1\over 2}\xi.\ee
Further finding the expressions for ${\partial \psi\over \partial t}$ and ${\partial \psi\over \partial r}$ from Eq. \eqref{49} and substituting these expressions in Eq. \eqref{8} together with the value of $v$ from Eq. \eqref{411} we obtain

\be da(\xi) = \left[{D\over (3\xi - 2\sqrt{\xi})}\right] d\xi,\ee
which on integration yields

\be\label{423} a(\xi) = {2\over 3} D \ln{\left(\sqrt\xi - {2\over 3}\right)} + \mathrm{constant}.\ee
With this form of $a$, one can explicitly express $\psi$ appearing in Eq. \eqref{49} in the form

\be\label{414} \psi = {2\over 3}D \ln{\left(\sqrt \xi - {2\over 3}\right)} + D\ln{r} +\mathrm{constant}.\ee
Again a combination of Eqs. \eqref{5} and \eqref{26} yields

\be {1\over r^2}\left({\partial \psi\over \partial r}\right)\Bigg|_{r = \mathrm{constant}} = {n\lambda\over r^3}.\ee
We can obtain ${\partial \psi\over \partial r}$ from Eq. \eqref{414}, which, when used in the above relation, gives

\be\label{415} \eta = {\eta_1\over \left(2-{3\sqrt \xi}\right)},\ee
where $\eta_1$ is a constant, gven by $\eta_1 = {2 D\over \lambda}$. So ultimately we obtain the complete solution for this case with $\theta = 3$, $C = \mathrm{constant}$, and $V \ne {2\over 3}$. All the dimensionless quantities are also known, viz.$V(\xi)$ from Eq. \eqref{410}, $\phi(\xi)$ from Eq. \eqref{412}, $\eta(\xi)$ from Eq. \eqref{415}, and $Q = 0$, since the thermodynamic pressure $p = 0$. This solution has already been obtained by Chi (see solution (9) of \cite{3}), but here in Eq. \eqref{48}, one can choose $V$ arbitrarily as a function of $\xi$ or a constant, so that $a(\xi)$ can be found from Eq. \eqref{48} and thus $\psi$ is known explicitly from Eq. \eqref{49}. Thus other quantities can also be easily obtained, and hence all possible solutions for different values of $V$ can be generated, whereas Chi\cite{3} could obtain only a few special solutions. We shall now discuss
some of the solutions with a special choice of $V$. Since the solution for $V = 1$ has already been obtained by Chi (see Eq. (5) of \cite{3}), we shall consider in the following examples some other values of $V \ne 1$.\\

\noindent
\textbf{Example 1}. $V =2$:\\
With this value of $V$, it is possible to find $a(\xi)$ from Eq. \eqref{48} as

\be \label{416} a(\xi) = \left({D\over 2}\right)\ln{\xi} + \mathrm{constant}.\ee
With the help of Eq. \eqref{416}, $\psi$ can be found from Eq. \eqref{49} as

\be \label{417} \psi = \left({D\over 2}\right)\ln{\xi} + D\ln{r} + \mathrm{constant}.\ee
So now $\eta$ can be found out from Eqs. \eqref{5}, \eqref{26}, and \eqref{417} as

\be\label{418} \eta = -{1\over 2}\left({D\over 2}\right) = \eta_0,\ee
where $\eta_0$ is a constant. We therefore obtain $\phi$ from Eq. \eqref{32} as

\be \label{419} \phi = \left({C\over G\lambda}\right)\xi = \phi_0\xi,\ee
where, $\phi_0 = {C\over G\lambda}$ is a constant. For $V = 2$, it is found from Eq. \eqref{5} that

\be v = r_t = 2\left({r\over t}\right),\ee
and a second differentiation with respect to time coordinate at constant $\psi$ yields

\be v_t = r_{tt} = 2\left({r\over t^2}\right).\ee
Further using \eqref{5} in the relation \eqref{49} one also finds

\be g = \phi_0\left({r\over t^2}\right) \xi.\ee
With the help of the above two equations one can directly integrate Eq. \eqref{27} and get the solution for $p$ in the form

\be\label{420} p = 2\lambda{\eta_0\over t^2 r} - c\lambda{\eta_0\over 4r^4} + T(t),\ee
where $T(t)$ is an arbitrary function of $t$. So all the quantities are known with this special choice of $V = 2$. This is a new solution. It is to be noted that to find $p$, Eq. \eqref{27}, instead of
Eq. \eqref{24}, has been utilized, because it gives directly the solution for $p$ as a function of $r$ and $t$.\\

\noindent
\textbf{Example 2}. $V = \xi$:\\
In this example, instead of considering $V$ to be a constant, we have chosen $V$ as a simple function
of $\xi$. The solution can be obtained by the above procedure. We obtain $a(\xi)$ from Eq. \eqref{48} as

\be a(\xi) = {\left({D\over 3}\right)}\ln{(3\xi -2)}.\ee
We can obtain $\psi$ from Eqs. \eqref{49} and \eqref{42}) as

\be\label{422} \psi = \left({D\over 3}\right) \ln{(3\xi - 2)} + D\ln{r} +\mathrm{constant}.\ee
We can obtain $\eta$ from Eqs. \eqref{5}, \eqref{26}, and \eqref{422} as

\be\label{423} \eta = {\eta_0\over(2 - 3 \xi)},\ee
where $\eta_0 = {2D\over \lambda}$ is a constant quantity. We also find $\phi$ from Eq. \eqref{32} as

\be\label{424} \phi = \phi_0~\xi.\ee
The pressure $p$ can be obtained again in the same way as in the previous example. Also, $\rho$, $g$, and $v$ can be known from the value of $\xi$ given in Eq. \eqref{6}. All these give another set of new solutions.

\section{Solution for $\theta \ne 3$:}

We have already seen that if $\theta \ne 0$, the variable $r$, which is actually a function of the new variables $\psi$ and $t$, can be expressed in view of \eqref{31} and \eqref{32} in the form

\be \label{51} r = a(\psi)~B(\xi), \ee
where $a(\psi)$ and $B(\xi)$ are functions of $\psi$ and $\xi$, respectively. Equations \eqref{5} and \eqref{26} together give a relation like

\be \label{52} \rho = {1\over r^2 r_{\psi}} = {\lambda\eta\over r^\theta},\ee
so that one can write

\be r_{\psi} = {r^{(\theta -2)}\over \lambda\eta}.\ee
Again from Eq. \eqref{5} we know

\be v = r_t = \left({r\over t}\right)V,\ee
and as such one gets

\be {r_t\over r_{\psi}} = {\partial\psi\over \partial t}\Big|_{r = \mathrm{constant}} = -{\lambda \eta V\over t r^{(\theta -3)}},\ee
from which it immediately follows that

\be \label{53} {\partial \psi\over \partial \xi}\Big|_{r = \mathrm{constant}} = {\partial \psi\over \partial \left({G\lambda t^2\over r^\theta}\right)}\Bigg|_{r = \mathrm{constant}} = - {\lambda\eta V\over 2\xi (\alpha\beta)^{\theta - 3}}.\ee
Again from Eq. \eqref{51} one gets

\be dr = \beta~\alpha_{\psi}~d\psi + \alpha ~\beta_{\xi} ~d\xi,\ee
which enables us to write

\be \label{54} {\partial \psi\over \partial \xi}\Big|_{r = \mathrm{constant}} = - {\alpha\over \alpha_{\psi}}\left({\beta_{\xi}\over \beta}\right).\ee
Comparing Eqs. \eqref{53} and \eqref{54}, we arrive at the following results:

\be\label{55} {\alpha_{\psi}\over \alpha} = {\alpha^{(\theta - 3)} \over H},\ee
and also,

\be\label{56} {\beta_{\xi}\over \beta} = {\lambda\over H}\left[{\eta V\over 2\xi \beta ^{(\theta -3)}}\right],\ee
where $H$ is a constant. Integrating Eq. \eqref{55}, one further gets for $\theta \ne 0$

\be\label{57} \alpha^{3 - \theta} = \left[{3 - \theta \over H}\right]\psi + \mathrm{constant}.\ee
Again Eq. \eqref{51}, in view of \eqref{6}, gives

\be r_{\psi} = {\beta \alpha_{\psi}\over \left[1 + {\theta\xi \beta_{\xi}\over \beta}\right]}.\ee
We use this expression for $r_{\psi}$, in \eqref{52} and further utilizing \eqref{55} we obtain the following relation

\be\label{58} {\beta^{(\theta - 3)}\over \eta\left[1 + {\theta\xi \beta_{\xi}\over \beta}\right]} = {\lambda\over H}.\ee
Equations \eqref{56} and \eqref{58} together now yield

\be\label{59} V = {2\xi \beta_{\xi} \over \beta + \theta\xi \beta_{\xi}}.\ee
Also with the help of Eq. \eqref{51}, Eq. \eqref{32} can be written in the form

\be {C(\psi)\over \alpha^{(\theta -3)}} + G \lambda \beta ^{(\theta - 3)}\phi ~\xi^{-1}.\ee
Since the left-hand side of the above equation is a function of $\psi$ and the right-hand side is a function of $\xi$ the equation holds only if both sides are equal to a constant, that is

\be\label{510} C(\psi)  = E\alpha^{(3 - \theta)},\ee
and
\be\label{511} \phi(\xi) = \left({E\over G\lambda}\right)\xi \beta^{(\theta - 3)},\ee
where $E$ is a constant. Now the complete solutions for the $\theta \ne 3$ case are obtained from the set of equations \eqref{57} - \eqref{511}. If, for example, $V$ is given as a function of $\xi$, then from Eq. \eqref{59} $\beta$ can be found as a function of $\xi$ and as such $\beta$ can be written explicitly as a function of $r$ and $t$. Hence $\eta$ and $\phi$ can be obtained from Eqs. \eqref{58} and \eqref{511}, respectively. As $\beta$ is now known, $\alpha$ can also be found as a function of $r$ and $t$ from Eq. \eqref{51}. So from Eq. \eqref{57}, $\psi$ can be determined as a function of $r$ and $t$. All the quantities being known, it is now easy to find the solution from the pressure $p$ from Eq. \eqref{27}. At this point we proceed to find the solutions identical with those appearing in the case of a Friedmann universe with zero cosmological constant, spatially uniform density distribution $[p = p(t)]$, and $p = 0$. We define the function $R(t)$ as (see Chi \cite{3})

\be \label{512} {v\over r} = {V\over t} = {\dot R\over R},\ee
where an overdot represents the time derivative. In Eq. \eqref{512}, $R(t)$ is analogous to the scale factor of the Friedmann universe and is dependent on time alone. Therefore, $V$ must be either a constant or a function of time alone. In the case $V = V(t)$, and $\xi$ is a function of time only, which implies $\theta = 0$ from the definition of $\xi$. Again spatially, uniform density distribution is a necessary condition for the Friedmann universe. Equation \eqref{5} implies that $\rho = \rho(t)$ is satisfied either for $\theta = 0$, or for $\eta(\xi) \simeq \xi^{-1}$. So in order to deal with the Friedmann model we have to consider either $V =$ constant, and $\eta \simeq \xi^{-1}$, or $\theta = 0$. The second case, that is, $\theta = 0$, however, gives only a limiting value for $\xi = \xi(t)$.\\

\noindent
\textbf{A. Case 1}. $\theta \ne 3,~ V = V_0 (\mathrm{constant})$, and $\eta = \eta_0 \xi^{-1}$:\\
Under these conditions, Eq. \eqref{59} can be easily integrated and one gets

\be \label{513} \beta ^{(2-\theta V_0)\over V_0} = A \xi,\ee
where, $A$ is the constant of integration. Since $\eta = \eta_0 \xi^{-1}$, Eq. \eqref{58} can now be easily simplified by using Eq. \eqref{513} to yield

\be \beta ^{(2- 3 V_0)\over V_0} = {\left(A\lambda \eta_0\over H\right)}\left[(2-\theta V_0)\over 2\right].\ee
Since the right-hand side of the above equation is a constant, it is satisfied only if the left-hand side is also equal to a constant, that is, when

\be {(2- 3V_0)\over V_0} = 0.\ee
This immediately gives us

\be \label{514} V = V_0 = {2\over 3},\ee
which implies, in view of Eq. \eqref{5},

\be \label{515} v = {2\over 3}\left({r\over t}\right).\ee
From Eq. \eqref{5} one can find the value of $\rho$ by using Eq. \eqref{6} and it is given by

\be\label{516} \rho = {\eta_0\over Gt^2}.\ee
Combining Eqs. \eqref{513} and \eqref{514} one gets

\be \xi \beta^{(\theta - 3)} = {1\over A}.\ee
Applying the above result in Eq. \eqref{511} it is not difficult to find that

\be \label{517} \phi = \phi_0,\ee
where, $\phi_0 = {E\over AG\lambda}$, $E$ being a constant quantity. Therefore from Eq. \eqref{5} one finds

\be \label{518} g = \phi_0\left({r\over t^2}\right).\ee
Since $v = r_t$ is known from Eq. \eqref{515} and also, since $g$ has
already been given by Eq. \eqref{518}, it is now possible to solve easily Eq. \eqref{27} for $p = 0$ to obtain

\be \label{519} \phi_0 = -{2\over 9},\ee
which, when applied in Eq. \eqref{518}, finally yields

\be \label{520} g = -{2\over 9}\left({r\over t^2}\right).\ee
Now with $V = {2\over 3}$, Eq. \eqref{512} becomes

\be \label{521} {\dot R\over R} = {2\over 3t},\ee
and it is thus possible to express the fluid density $\rho$ in the form

\be \label{522} \rho = {3\over 8\pi G} {\left(\dot R\over R\right)}^2,\ee
by suitably choosing the constant $\eta_0$. This is simply the Friedmann differential equation with zero cosmological constant and zero spatial curvature $(k = 0)$. It is a case of the Einstein-de Sitter universe as given by Chi \cite{3} (see the solution (13) of Chi \cite{3}).\\

\noindent
\textbf{B. Case 2}. $\theta = 0$:\\
In this case it is obvious from Eq. \eqref{5} that $\rho = \rho(t) = \lambda \eta$. Now since $\xi$ is a function of $t$ alone, so undoubtedly $\eta(\xi)$ is also a function of $t$ only. Hence in view of Eqs. \eqref{511}, \eqref{6}, and \eqref{5}, one gets

\be \label{523} g = E \beta^{-3} r.\ee
Again, since $\beta_t = \beta_{\xi} \xi_t$ one can find $\xi_t$ from Eq. \eqref{6} and hence $\xi \beta_{\xi}$ can be obtained as

\be \xi\beta_{\xi} = {t\beta_t\over 2},\ee
which, being used in Eq. \eqref{59}, yields

\be \label{524} V = {t\beta_t\over \beta}.\ee
Using Eqs. \eqref{512} and \eqref{524} we find

\be \label{525} {\dot R\over R} = {\dot \beta\over \beta},\ee
which on integration yields

\be \label{526} \beta = \beta_0 R.\ee
Here, $\beta_0$ is the constant of integration, and $\eta$ can be found from Eq. \eqref{58}, which, again being applied in the equation \eqref{5}, gives

\be \rho~\beta^3 = H,\ee
and the above equation, together with Eq. \eqref{526} finally expresses the density $\rho$ in the form

\be \label{527} \rho = {\rho_0\over R^3},\ee
here $\rho_0$ is constant. This is a well-known result for the Friedmann dust universe. Again from Eqs. \eqref{5} and \eqref{511} we have

\be \label{528} g = E~\beta^{-3} r.\ee
Also using Eqs. \eqref{5} and \eqref{524} one finds

\be r_t = {r\beta_t\over \beta},\ee
from which $r_{tt}$ can be easily found. With the form of $r_{tt}$ being calculated from the above expression and using the result of Eqs. \eqref{528} and \eqref{25}, it is possible now to solve Eq.
\eqref{27} for dust, that is, $p = 0$. We thus obtain

\be \beta^2\beta_{tt} = E.\ee
Integrating the above equation we find

\be {\dot\beta\over \beta} = \pm \left({1\over \beta}\right)\left[-{E\over \beta} + E_1\right]^{1\over 2},\ee
where $E_1$ is a constant of integration. Combining the above equation with Eq. \eqref{525} and using the result of Eq. \eqref{526} we find

\be \dot R^2 = \left(-{E\over \beta_0^3 R} + {E_1\over \beta_0^2}\right).\ee
Assuming $(-{E\beta_0^{-3}})$ to be positive everywhere, i.e., $(-{E\beta_0^{-3}} =  m^2)$, say, where $m$ is any real number, and also writing ${E_1 \beta_0^{-2}} = -k$ we find

\be \dot R^2 = {m^2\over R} - k,\ee
where $k$ may be positive, negative or zero. The above equation leads to the result

\be {\dot R^2\over R^2} + {k\over R^2} = {m^2\over R^3}.\ee
Using Eq. \eqref{527} in the above relation, we finally obtain:

\be \label{529} {\dot R^2\over R^2} + {k\over R^2} = \left({m^2\over \rho_0}\right)~\rho.\ee
Identifying ${m^2\over \rho_0}$ with ${8\pi G \over 3}$, it is now possible to recognize the above equation to be the differential equation for the Friedmann universe with zero cosmological constant and non vanishing spatial curvature ($k = \pm 1$), the $\left(^0_0\right)$ equation of Einstein. So the integration constant $E_1$ is associated with the space curvature $k$ of the Friedmann universe. The vanishing of the spatial curvature $(k = 0)$ leads to $R \sim t^{2\over 3}$ or $\beta \sim t^{2\over 3}$ and $\eta = \eta_0~\xi^{-1}$, which is exactly the Einstein--de Sitter solution obtained by Chi, \cite{3} except for the fact that here $\theta$ assumes only a fixed value zero.\\

We shall now proceed to present more solutions with different choices of $V$ for a perfect fluid ($p \ne 0$) and for a spatially uniform density [$\rho = \rho(t)$] with $\theta = 0$.\\

\noindent
\textbf{1. Case 2a.} $V = V_0 = \mathrm{constant}, \theta = 0$:\\
In view of Eq. \eqref{5} we have

\be\label{530} v = V_0\left({r\over t}\right).\ee
Equation \eqref{6} yields

\be\label{531} \xi = G\lambda t^2.\ee
Equation \eqref{59} can be easily integrated to yield

\be\label{532} \beta = (\beta_0 ~\xi)^{V_0\over 2},\ee
where $\beta_0$ is a constant of integration. We can obtain $\eta$ from Eqs. \eqref{52} and \eqref{532} as

\be\label{533} \eta = \eta_0~\xi^{-{3V_0\over 2}},\ee
where $\eta_0 = (H\lambda^{-1})\beta_0^{-{3V_0\over 2}}$ is a constant. So $\rho$ can be found from Eqs. \eqref{5}, \eqref{531}, and \eqref{533} as

\be\label{534} \rho = \lambda \eta_0(G\lambda)^{-{3V_0\over 2}} t^{-3V_0}.\ee
We can now obtain $\phi$ from Eq. \eqref{511} as

\be \label{535} \phi = \left({E\over G\lambda}\right)\xi \beta^{-3}.\ee
Equations \eqref{5}, \eqref{531}, and \eqref{535} together give

\be\label{536}  g = E \beta^{-3} r.\ee
With all these above expressions for $v$, $\rho$ , and $g$, presented in Eqs. \eqref{530}, \eqref{534} and \eqref{536} respectively, Eq. \eqref{27} takes the form

\be \label{537} {\partial p\over\partial r} = \lambda\eta_0(G\lambda)^{-{3\over 2}V_0} ~t^{-3V_0}\left[{Er\over (\beta_0 G\lambda)^{{3\over 2}V_0}t^{3V_0}} - (V_0^2 - V_0){r\over t^2} \right].\ee
If we now consider the case $V =1$, with the help of the set of equations \eqref{530} - \eqref{537}, we arrive at the following set of solutions:

\be\begin{split}\label{538} &v = {r\over t},\hspace{0.9 in} \rho = \lambda \eta_0 (G\lambda)^{-{3\over 2}} t^{-3},\\&
g = Er(\beta_0~\xi)^{-{3\over 2}}, \hspace{0.32 in}p = \left({E\lambda\eta_0\over 2 \beta^{3\over 2}(G\lambda)^3}\right)\left({r^2\over t^6}\right) + T(t).\end{split}\ee
Here the constant of integration $T(t)$ must be equal to zero, since $\xi = \xi(t)$ and $p = \lambda r^2 t^{-2} Q(\xi)$ [from Eq. \eqref{5}]. Finding ${\partial \over \partial t}(p\rho^{-\gamma})$ and ${\partial \over \partial r}(p\rho^{-\gamma})$ and using Eqs. \eqref{3} and \eqref{538} one finds immediately that $\gamma = {4\over 3}$. So the above uniform density solution \eqref{538}, with $\theta = 0$ and $V = 1$, is valid only for $\gamma = {4\over 3}$ polytrope.\\ 

\noindent
Next, let us consider $V_0 = 2$ using the set of relations \eqref{530} - \eqref{537}, we can again obtain the following solutions:

\be \begin{split}& v = 2{r\over t}, \hspace{1.0 in}\rho = {H\over \beta_0^3}\left({1\over \xi^3}\right) \propto {1\over t^6},\\&g = {Er\over \beta_0^3 G\lambda t^2 \xi^2} \propto {r\over t^6}, \hspace{0.26 in}p = {2\over 3 a_2}{r^3\over t^8} - {a_1\over 2a_2}{r^2\over t^6} + \mathrm{constant}.
\end{split}\ee
In the above $a_1$ and $a_2$ are two constants given by

\be a_1 = {E\over (\beta_0 G\lambda)^3},~~~~~a_2 = {(\beta_0 G\lambda)^3\over H}.\ee
This is again another new set of solutions.\\

\noindent
\textbf{2. Case2b.} $V=2\xi, \theta = 0$:\\
In this case instead of considering $V$to be a constant, we have considered a simple functional relation between $V$ and $\xi$. It is possible to get the relevant solutions following the same procedure, and these are given by

\be \label{540} v = 2 G\lambda r t, ~~~\rho = H \beta_0^{-3} e^{-3G\lambda t^2},~~~g = E \beta_0^{-3} re^{-3G\lambda t^2}, ~~~p = F(t) r^2.\ee
The solutions \eqref{540} are also another new set of solutions obtained by our procedure. So in this way we can now choose any arbitrary form for $V$ to generate variety of possible solutions.

\section{Conclusion:}

In effect we have systematically derived a complete set of solutions for the hydrodynamic self-similar cosmological models with the help of a new set of independent variables $(\psi, t)$ in place of the old variables. Finally all the physical quantities have been expressed in terms of the old variables $(r, t)$. Some of these solutions are in agreement with those previously obtained by Chi in an ad hoc manner. We have further obtained the Newtonian analogs of all homogeneous isotropic Friedmann dust universes with spatial curvature $k = 0 ~\mathrm{as~ well~ as}~ \pm 1$.\\

\noindent
\textbf{Acknowledgement:}\\
The work has been carried out under the financial support of the ``University Grants Commission", India.

\end{document}